\documentclass[nocompress]{spie}

\usepackage{amsmath,amsfonts,amssymb}
\usepackage{graphicx}
\usepackage[colorlinks=true, allcolors=blue]{hyperref}

\usepackage{subcaption}
\usepackage{multirow}
\usepackage{bm}
\usepackage[labelfont=bf, skip=8pt]{caption}
\usepackage[indLines=false]{algpseudocodex}

\usepackage{fancyhdr}
\usepackage{booktabs}

\newcommand{\C}{\mathcal{C}}

\newcommand{\A}{\mathbb{A}}
\newcommand{\Hc}{\mathbb{H}_{\C}}
\newcommand{\ext}{{\text{ext}}}

\def\PAnumber{AFRL-2026-3596}

\fancypagestyle{empty}{%
  \fancyhf{} 
  \fancyfoot[C]{Approved for public release; distribution is unlimited.  Public Affairs release approval \# \PAnumber .}
  
}

\title{Outpainting: spatially extending aero-optic phase screens}

\author[a]{Jeffrey W. Utley}
\author[a]{Gregery T. Buzzard}
\author[b]{Charles A. Bouman}
\author[c]{Matthew R. Kemnetz}
\affil[a]{Department of Mathematics, Purdue University, West Lafayette, Indiana 47907, USA}
\affil[b]{Departments of Electrical and Computer Engineering, and Biomedical Engineering, Purdue University,  West Lafayette, Indiana 47907, USA}
\affil[c]{Department of Engineering Physics, Air Force Institute of Technology, Wright-Patterson AFB, OH 45433, USA}
\authorinfo{Further author information: (Send correspondence to J.W.U.)\\J.W.U.: E-mail: utleyj@purdue.edu}

\begin{document} 
\maketitle

\begin{abstract}
    Aero-optic effects distort light wave propagation near a high-speed aircraft, thereby degrading performance in airborne imaging and communication systems. Measuring aero-optic data through experiment is costly and the resulting data often has a limited spatial size. Further, alternative methods for simulating this data, including computational fluid dynamics and conventional phase screen generation algorithms (e.g., boiling flow), face drawbacks such as large computation time or inaccurate statistics. More recently, data-driven algorithms have been proposed that can synthesize data that matches relevant statistics of measured aero-optic data. However, these methods cannot spatially extend aero-optic data. In this paper, we introduce ReVAR-ext (Re-whitened Vector AutoRegression-extender), an algorithm that builds on an existing data-driven approach, ReVAR, to spatially extend measured aero-optic data (a process called outpainting) and match the spatial and temporal correlations of the measured data. ReVAR-ext generalizes the generation process of ReVAR by combining multiple sets of synthetic data with the input measured data. This approach generates multiple fixed-sized synthetic images, each of which overlaps with the input data, and then stitches them together. When paired with ReVAR, the ReVAR-ext algorithm can generate aero-optic data with arbitrary temporal duration and arbitrary spatial size. Our experiments show that extended data generated by ReVAR-ext closely matches the temporal power spectrum of two measured aero-optic data sets. Further, the extended data approximately matches the spatial autocorrelation, with reduced accuracy at large spatial lags and at vertical lags.
\end{abstract}


\section{INTRODUCTION}
\label{sec:intro}
As an aircraft flies through the atmosphere at high speeds, aerodynamic turbulence near the aircraft distorts light wave propagation. Specifically, light waves propagating near the aircraft pass through a region of aerodynamic turbulence, resulting in phase aberrations called aero-optic effects \cite{Visbal, Kalensky, WangPhysicsComputation}. Measuring these phase aberrations through experiment results in a time-series of images called phase screens \cite{Kemnetz, Geary, Holmes}. However, these measured phase screens are often restricted to a small spatial size \cite{KemnetzAnalysis}. This limits the data available for testing systems and developing mitigation algorithms that must account for aero-optic effects.

Due to these limitations of measured aero-optic data, simulations have been used to generate synthetic data. For example, Computational Fluid Dynamics (CFD) simulations can be used to generate synthetic aero-optic phase screens \cite{WangAero-Optics, WangComputation, Porter}. However, as discussed in more detail in Ref.~\citenum{UtleyBoiling2}, there is a trade-off between data quality and computational efficiency for CFD algorithms.

Conventional phase screen generation methods, such as boiling flow, can also generate time-series of arbitrarily sized images. The boiling flow algorithm \cite{Srinath} uses well-established theory for atmospheric turbulence \cite{Schmidt, Tatarski, Taylor, PoyneerExperimental} to generate synthetic phase screen data. This method is computationally efficient \cite{Srinath} and can generate data with arbitrary spatial size. However, it relies on statistical assumptions, including spatial homogeneity, that do not always apply to aero-optic phase aberrations \cite{Vogel, UtleyBoiling, UtleyBoiling2, Siegenthaler}. Although Utley \textit{et al.}\cite{UtleyBoiling, UtleyBoiling2} adapted the boiling flow algorithm to generate synthetic aero-optic phase screens, the resulting synthetic data could not simultaneously match both the spatial and temporal correlations of measured aero-optic data.

Recently, data-driven methods have been developed to generate synthetic aero-optic data that matches the statistics of measured data. For example, Utley \textit{et al.}\cite{Utley1, Utley2}, Faghihi \textit{et al.}\cite{Faghihi}, and Vogel \textit{et al.}\cite{Vogel} have used linear autoregressive models that train on measured aero-optic data and can then generate arbitrary-duration synthetic time-series data. These methods have been shown to closely match relevant spatial and temporal statistics of measured aero-optic data\cite{Faghihi, Utley2}. However, they do not have a mechanism to spatially extend measured aero-optic data.

In this paper, we introduce ReVAR-ext (Re-whitened Vector AutoRegression-extender), an extension of the recently introduced ReVAR algorithm \cite{Utley2, ReVAR_Code} that spatially extends measured aero-optic phase screens outside of their limited domain (a process called outpainting). ReVAR-ext takes an input time-series of phase screens and then extends each input phase screen to a larger size. This algorithm first generates multiple sets of synthetic, fixed-sized phase screens, each of which overlaps with the input phase screens, and then stitches these synthetic phase screens to the input phase screens along the overlap. Combining ReVAR with ReVAR-ext can then generate time-series of phase screens with arbitrary spatial size and arbitrary temporal duration. Our experiments show that data generated by ReVAR-ext closely matches the temporal power spectrum (TPS) of two measured aero-optic data sets within 5\% normalized root mean square error (NRMSE) and approximately matches the spatial autocorrelation within $\approx$17\% NRMSE. Further, ReVAR-ext matches the TPS of synthetic data sets within 3\% NRMSE and the spatial autocorrelation within 18\% NRMSE.

This paper builds on previous work from Refs.~\citenum{Utley1, Utley2}.

\begin{figure}[h]
    \centering
    \includegraphics[width=0.8\textwidth]{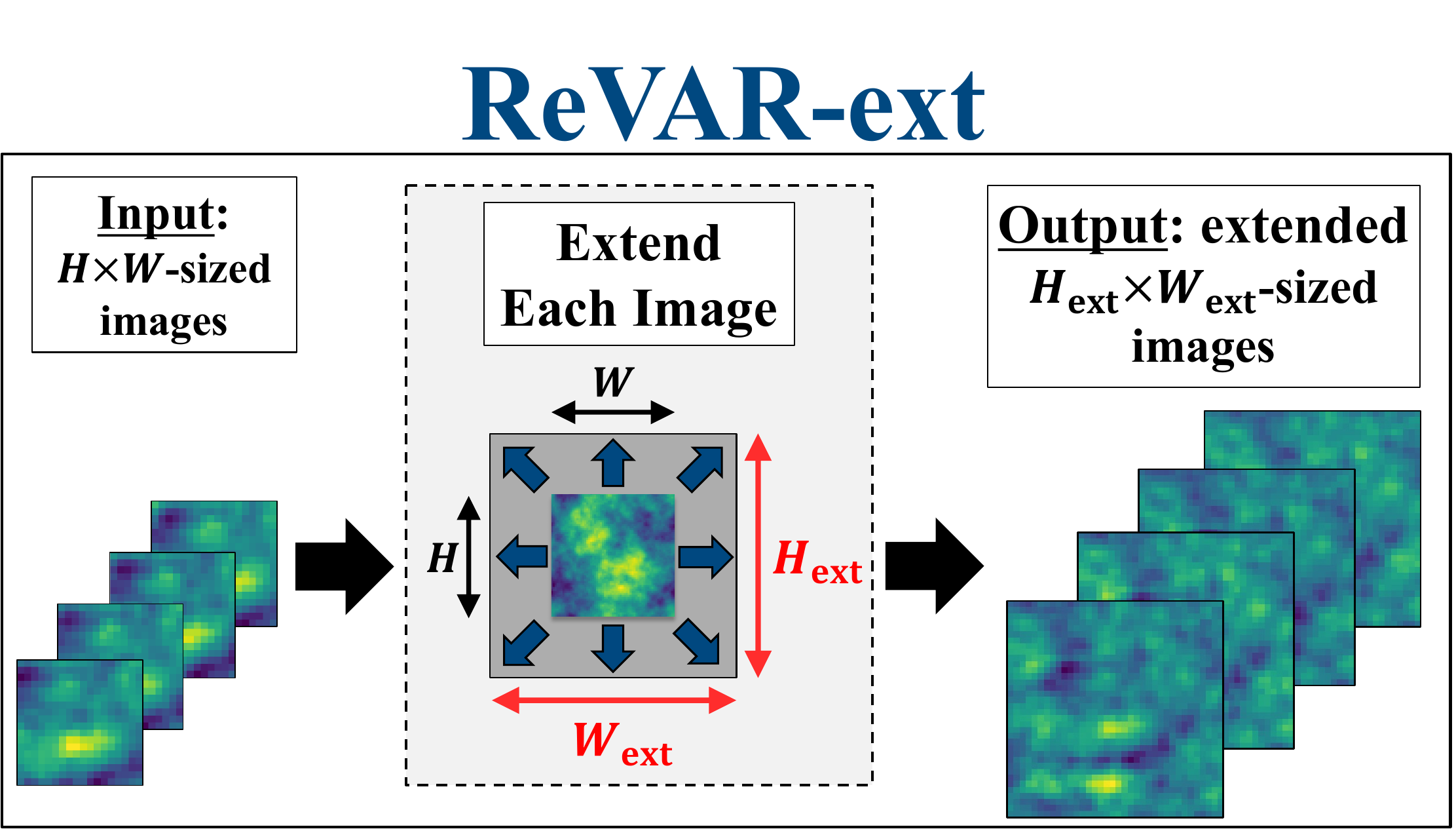}
    \includegraphics[width=0.8\textwidth]{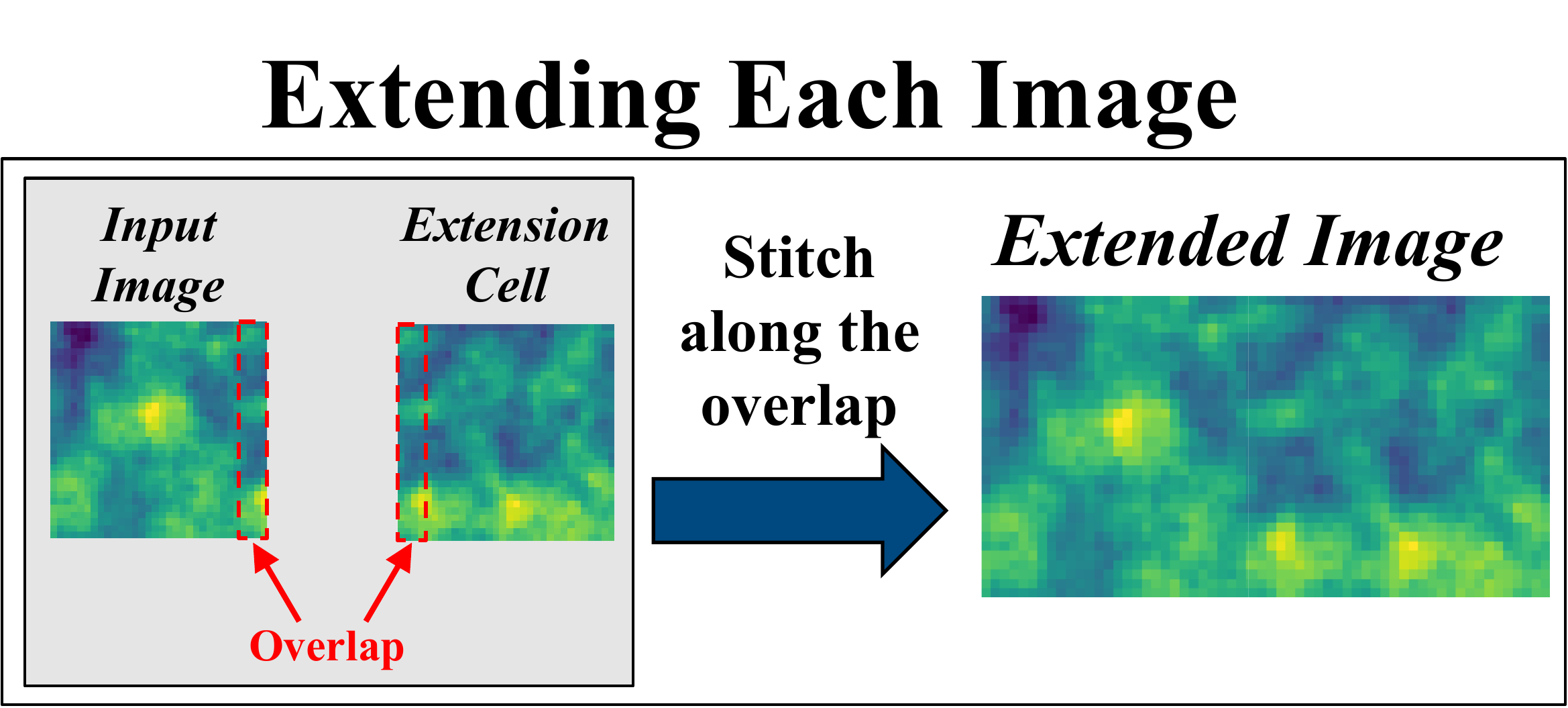}
    \caption{Outline of the ReVAR-ext algorithm. {\bf Top:} Input and output of ReVAR-ext. This algorithm takes in an input time-series of images with a fixed size $H\times W$. ReVAR-ext then places each input $H\times W$-sized image inside a larger, $H_\ext\times W_\ext$-sized spatial domain and extends the images to this larger $H_\ext\times W_\ext$ size. After applying this process at each time-step, ReVAR-ext outputs a time-series of extended, $H_\ext\times W_\ext$-sized images. {\bf Bottom:} Methodology for extending images. For each input image, we generate a synthetic image called an extension cell that overlaps with the input image. We then stitch the extension cell to the input image along this overlap to create a larger, extended image.}
    \label{fig: Figure 1}
\end{figure}

\section{ReVAR-ext: ALGORITHM DESIGN}\label{s: ReVAR-ext}
The ReVAR-ext algorithm spatially extends a time-series of (single-channel) images outside of their limited, fixed spatial domain. This method builds on the existing ReVAR algorithm \cite{Utley2, ReVAR_Code}, which generates arbitrary-duration time-series of images that match the spatial and temporal correlations of measured data. Synthetic images generated by ReVAR have the same size as the measured images, so ReVAR-ext generalizes the generation process by combining multiple sets of synthetic images with the measured images. This combination of synthetic images with measured images creates a time-series of larger images.

Figure~\ref{fig: Figure 1} illustrates the core methodology of ReVAR-ext. Given an input time-series of images with a fixed size $H\times W$, ReVAR-ext can spatially extend the images to an arbitrary size $H_\ext \times W_\ext $. This algorithm achieves this larger $H_\ext \times W_\ext $ size by centering the input $H\times W$-sized images inside of the $H_\ext \times W_\ext $ spatial domain and then extending on each side of the input images. For each of these extensions, we generate a time-series of new $H\times W$-sized images that overlap with the input images; we call these newly generated images \textit{extension cells}. We then stitch the extension cells to the input images along this overlap, as shown in Fig.~\ref{fig: Figure 1}, bottom. 

Because ReVAR-ext is built on top of ReVAR, we can derive the parameters of ReVAR-ext from a trained ReVAR model and do not require any additional parameter estimation from measured data.

\subsection{ReVAR: Autoregressive Generation}\label{s: ReVAR Classical}
The ReVAR algorithm is described in detail in Ref.~\citenum{Utley2}. In this section, we highlight the key features of ReVAR used by ReVAR-ext and emphasize ReVAR's inherent restriction to a fixed image size.

ReVAR generates a synthetic time-series of images using a linear autoregressive model applied to the Principal Component Analysis (PCA) coefficients of the images:
\begin{align}\label{eq: Autoregressive Model}
    \tilde{X}_n = \hat{X}_n + \xi_n.
\end{align}
Here, $\tilde{X}_n$ contains the PCA coefficients of the image at time-index $n$, $\hat{X}_n$ is a linear predictor of $\tilde{X}_n$ given the time-history $(\tilde{X}_0,\dots,\tilde{X}_{n-1})$, and $\xi_n\sim \mathcal{N}(\bm{0}, R_\xi)$ is additive noise. The linear predictor used in this autoregressive model is given by
\begin{align}\label{eq: Long-Range Predictor}
    \hat{X}_n = \sum_{\ell=1}^{N_L}A_{X,\ell} \tilde{X}_{n-\ell} + \sum_{i=1}^{N_{LPF}} A_{Y,i} Y_{i,n-1},
\end{align}
where $A_{X,\ell}, A_{Y,i}$ are the weights of this linear prediction and 
\begin{align}\label{eq: Low-Pass Filters}
    Y_{i,n} = (1-\alpha_i)\: Y_{i,n-1}+\alpha_i\:\tilde{X}_n
\end{align}
are infinite impulse response low-pass filters of the PCA coefficient time-series with weights $\alpha_i$. After generating synthetic PCA coefficients $\tilde{X}_n$ through Eq.~\eqref{eq: Autoregressive Model}, ReVAR applies the following equation to generate synthetic images:
\begin{align}\label{eq: Synthetic Images}
    X_n^{\text{syn}} &= E \tilde{X}_n.
\end{align}
Here, the matrix $E$ contains the principal component basis and converts from PCA coefficients to pixel values.

Although this linear autoregressive model can generate arbitrarily many time-steps of synthetic data through recursive application, the spatial dimensionality of the generated synthetic data is restricted to the size of the training data. This restriction primarily arises from the use of a PCA basis \cite{Chatterjee, Berkooz}. The dimensions of the principal component matrix $E$ must always equal the spatial dimensionality of the training data, so the synthetic images $X_n^{\text{syn}}$ are restricted to this limited spatial dimensionality.

\subsection{ReVAR-FFBS: Generating Extension Cells}\label{s: ReVAR-FFBS}
Since ReVAR is limited to the same spatial dimensionality of the training data, we instead adapt its methodology to generate each $H\times W$-sized extension cell used in the ReVAR-ext algorithm. Here, we sample from the autoregressive model of Eq.~\eqref{eq: Autoregressive Model} to match some known pixel values from the input images. We accomplish this by relating this problem to a dynamic linear model with partial observations and applying the existing Forward Filtering, Backward Sampling (FFBS) algorithm \cite{Carter, Fruhwirth-Schnatter, Petris}. We call this method ReVAR-FFBS.

\begin{figure}[h]
    \centering
    \includegraphics[width=\textwidth]{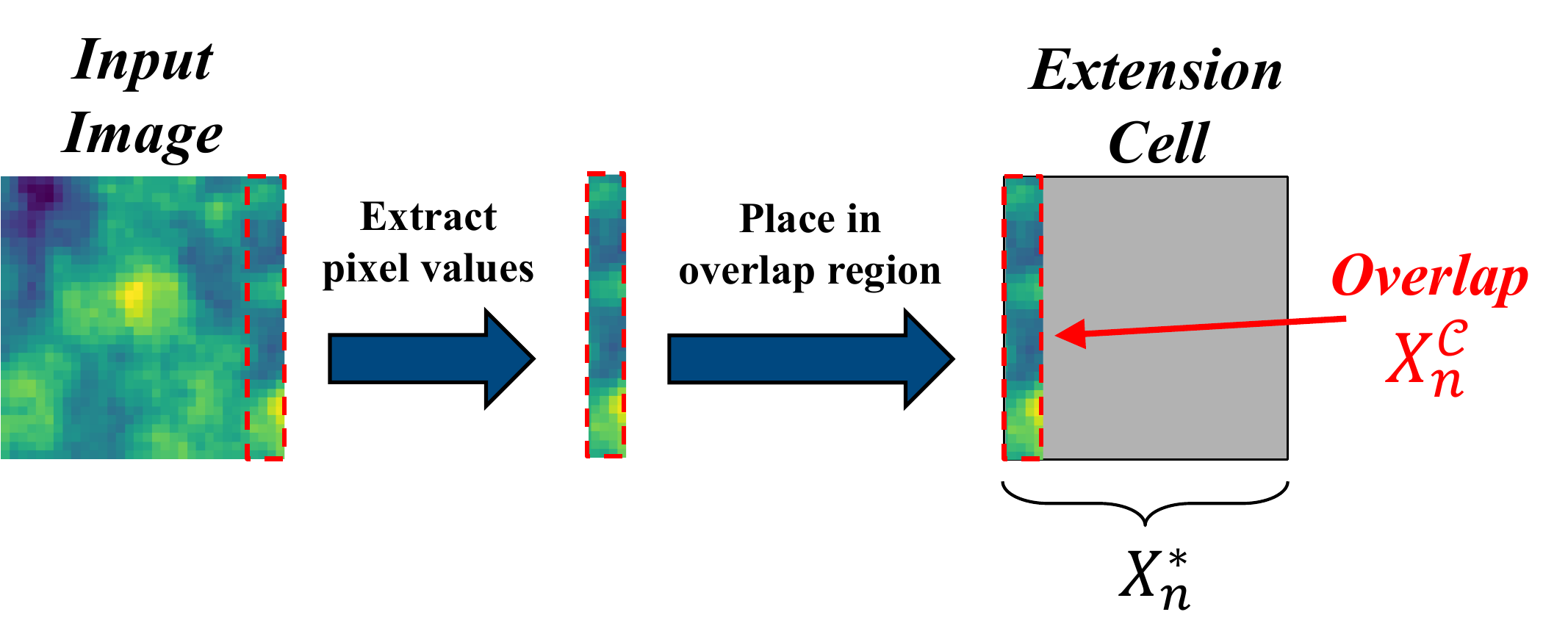}
    \caption{Example construction of the overlap-region and overlap-pixels for an extension cell. We take a set of pixel values from the edge of the input image at time $n$ and place them on the opposite edge of an extension cell. We call the pixel indices of this opposite edge the overlap-region and the set of pixel values the overlap-pixels; we denote these objects by $\C$ and $X_n^\C$, respectively. We then require that the extension cell at time $n$, which we denote $X_n^*$, matches these overlap-pixels $X_n^\C$ in the overlap-region $\C$. Extracting these pixels at all time-steps $n=0,\dots,N_T-1$ results in a time-series of overlap-pixels, $X_0^\C,\dots,X_{N_T-1}^\C$, and a time-series of extension cells, $X_0^*,\dots,X_{N_T-1}^*$.}
    \label{fig: Figure 2}
\end{figure}

Figure~\ref{fig: Figure 2} demonstrates how we extract, from the input images, the known pixel values that we require the extension cells to match. For each extension cell, we select (1) a set of pixels from the input images and (2) the pixel indices within the extension cell that should match these values. We call the pixel indices $\C$ the \textit{overlap-region} and the known values, $X_n^\C$ for $0 \leq n < N_T$, the \textit{overlap-pixels}, where $N_T$ is the number of time-steps of input data. We denote the extension cells by $X_n^*$ and require that they match these overlap-pixels $X_n^\C$.

To adapt ReVAR for extension cell generation, we sample from the autoregressive model of Eqs.~(\ref{eq: Autoregressive Model}-\ref{eq: Low-Pass Filters}) but add a constraint to ensure that the generated extension cells match the overlap-pixels. Following ReVAR's methodology, we first generate synthetic PCA coefficient vectors $\tilde{X}_n$ and then convert to images using the matrix $E$ from Eq.~\eqref{eq: Synthetic Images}. The restriction to $X_n^\C$ within the overlap-region $\C$ can then be expressed as the following \textit{constraint} that we enforce at all time-steps $n=0$ to $N_T-1$:
\begin{align}\label{eq: Constraint}
    E_\C\tilde{X}_n = X_n^\C.
\end{align}
The matrix $E_\C$ contains the set of rows of the PCA basis matrix $E$ corresponding to the overlap-region $\C$.

\paragraph{Dynamic linear model (DLM):} We can relate this model for extension cells to a dynamic linear model (DLM) \cite{Petris} by treating the AR process of Eqs.~(\ref{eq: Autoregressive Model}-\ref{eq: Low-Pass Filters}) as a state-transition equation and the time-series $X_n^\C$ as observations.

To convert the AR process to a state-transition equation, we create an augmented \textit{state vector} containing the AR process $\tilde{X}_n$, the lags $\tilde{X}_{n-\ell}$, and the low-pass filters $Y_{i,n}$:
\begin{align}\label{eq: State Vector}
    \tilde{\bm{X}}_n = \begin{bmatrix}
        \tilde{X}_n^T & \cdots & \tilde{X}_{n-N_L+1}^T & Y_{1,n}^T & \dots & Y_{N_{LPF},n}^T
    \end{bmatrix}^T.
\end{align}
We then write the autoregressive model of Eqs.~(\ref{eq: Autoregressive Model}-\ref{eq: Low-Pass Filters}) and the constraint of Eq.~\eqref{eq: Constraint} in terms of the augmented state vector $\tilde{\bm{X}}_n$:
\begin{align}\label{eq: State-Space Model}
    \tilde{\bm{X}}_n &= \A \tilde{\bm{X}}_{n-1} + \bm{\xi}_n, \\ \label{eq: Observation Model}
    X_n^\C &= \Hc \tilde{\bm{X}}_n.
\end{align}
We derive the following quantities from the state vector formulation of Eq.~\eqref{eq: State Vector}, ReVAR's linear autoregressive model of Eqs.~(\ref{eq: Autoregressive Model}-\ref{eq: Low-Pass Filters}), and the constraint of Eq.~\eqref{eq: Constraint}:
\begin{itemize}
    \item $\A$ is the state-transition model, which does not vary in time. We derive $\A$ from the prediction weights $A_{X,\ell}$, $A_{Y,i}$ and low-pass filter parameters $\alpha_i$ of ReVAR's autoregressive model.

    \item $\bm{\xi}_n$ is the process noise with covariance derived from $R_\xi$ and $\alpha_i$.
    
    \item $X_n^\C$ are the observations, which we assume to be deterministic (i.e., there is no observation noise).

    \item $\Hc$ is the observation model, which also does not vary in time. We derive $\Hc$ from $E_\C$.
\end{itemize}
This is a simple case of a DLM from Ref.~\citenum{Petris}.

\paragraph{Forward Filtering, Backward Sampling (FFBS):} We then generate time-series data $\tilde{\bm{X}}_0,\dots,\tilde{\bm{X}}_{N_T-1}$ that follows the dynamics of Eq.~\eqref{eq: State-Space Model} and matches the known data $X_0^\C,\dots,X_{N_T-1}^\C$ through Eq.~\eqref{eq: Observation Model}. This generation method allows us to use ReVAR's autoregressive model while also enforcing the constraint of Eq.~\eqref{eq: Constraint}.

We use the FFBS algorithm, as described in detail in Ref.~\citenum{Petris}, to generate this data. FFBS samples from the joint distribution of all states $\tilde{\bm{X}}_n$ given all observations $X_n^\C$. Letting $\tilde{\bm{X}}_{0:N_T}=(\tilde{\bm{X}}_0,\dots,\tilde{\bm{X}}_{N_T-1})$ (and similarly for $X_{0:N_T}^\C$), we sample from the conditional distribution $\tilde{\bm{X}}_{0:N_T}\;|\;X_{0:N_T}^\C$.

Sampling from the conditional distribution $\tilde{\bm{X}}_{0:N_T}\;|\;X_{0:N_T}^\C$ satisfies our requirements for generating extension cells. The time-series of observations $X^\C_{0:N_T}$ encodes the spatial and temporal correlations of the input data, so conditioning our data generation algorithm on $X^\C_{0:N_T}$ allows us to match these correlations. Further, since we sample the state vectors $\tilde{\bm{X}}_n$ given these observations, we ensure that this generated data satisfies the observation equation~\eqref{eq: Observation Model}.

After generating the time-series of state vectors $\tilde{\bm{X}}_{0:N_T}$ using FFBS, we convert from state vectors to extension cells. We extract the vector $\tilde{X}_n$ from $\tilde{\bm{X}}_n$ (following the augmented state vector formulation of Eq.~\eqref{eq: State Vector}) and convert to images by applying
\begin{align}
    X_n^* = E\;\tilde{X}_n
\end{align}
for each $n=0,\dots,N_T-1$. The resulting synthetic images $X_n^*$ then follow the dynamics of ReVAR's autoregressive model and satisfy the constraint of Eq.~\eqref{eq: Constraint}.

After generating a time-series of extension cells $X_0^*,\dots,X_{N_T-1}^*$, we stitch the extension cells to the input images. At each time $n$, we stitch $X_n^*$ to the input image at time $n$ along the overlap-region (as illustrated in Fig.~\ref{fig: Figure 1}, bottom).

\subsection{ReVAR-ext: Recursive Image Extensions}\label{s: ReVAR-ext: Recursive}
We apply the ReVAR-FFBS algorithm recursively to extend the input images from a fixed size $H\times W$ to an arbitrary size $H_\ext \times W_\ext$. Since the formulation of ReVAR-FFBS in Sec.~\ref{s: ReVAR-FFBS} leaves the overlap-region $\C$ arbitrary, it can be chosen to be any set of pixel indices within the $H\times W$ image size. ReVAR-ext thus uses multiple instantiations of ReVAR-FFBS, each with a unique overlap-region, to extend the input images on all four sides.

\begin{figure}[h]
    \centering
    \includegraphics[width=0.7\textwidth]{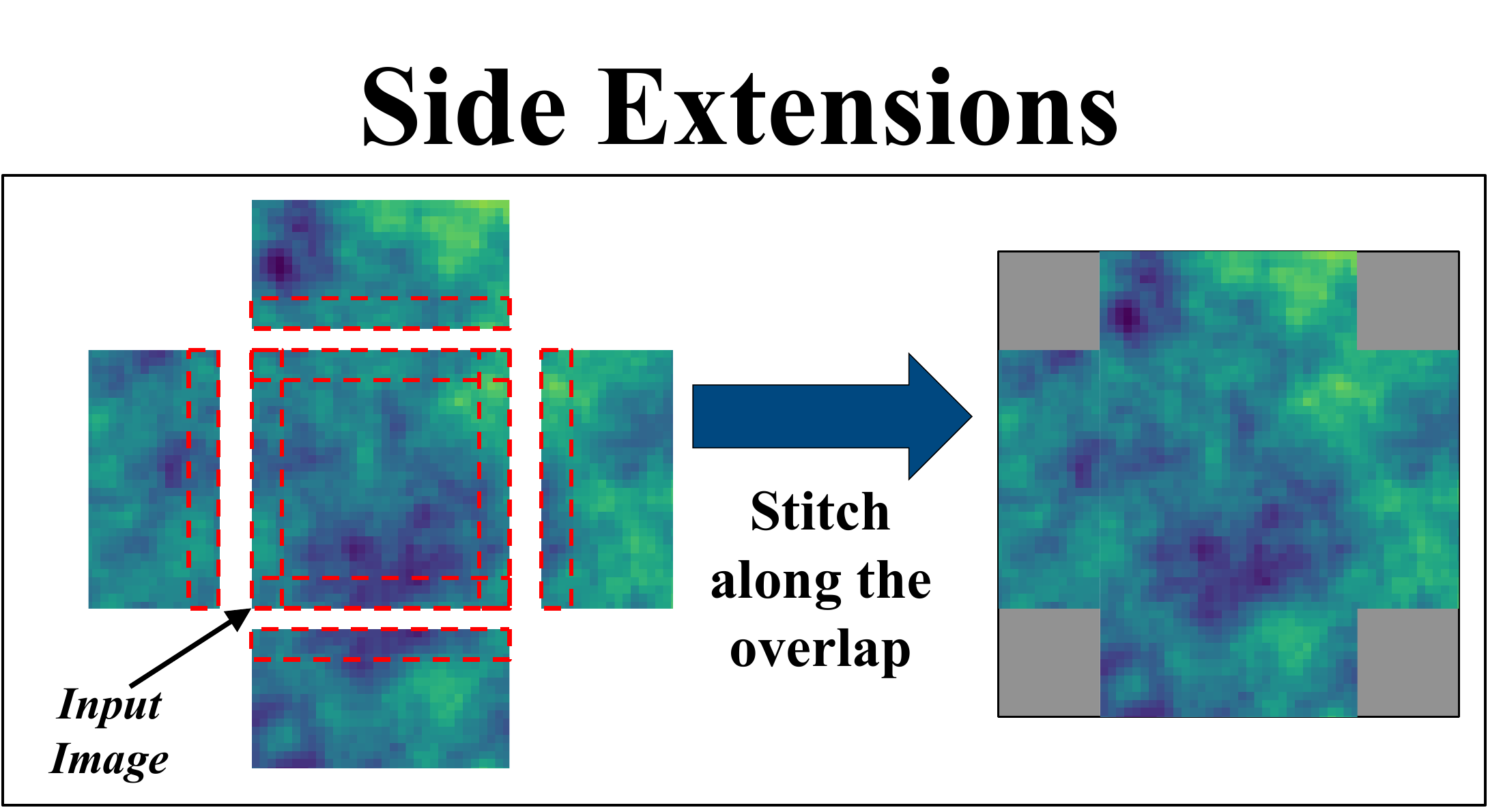}
    \includegraphics[width=0.7\textwidth]{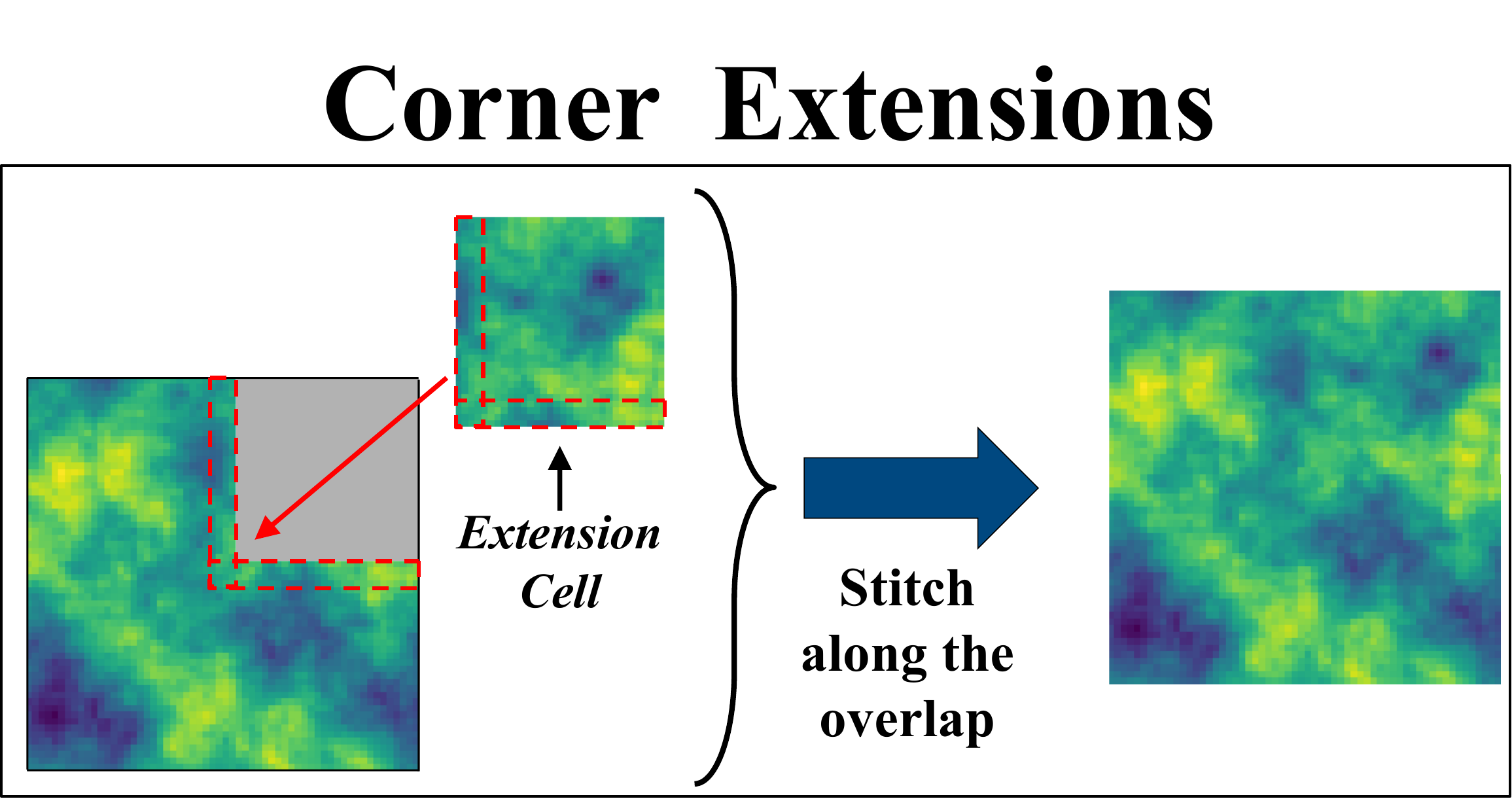}
    \caption{Overlap-regions used to extend the input images on all four sides. {\bf Top:} Overlap-regions used for side extensions. For each side, we take a vertical or horizontal strip containing a set of rows or columns at the edge of the input image. We then require that the opposite edge of each extension cell (i.e., the overlap-region) has the same pixel values as the chosen edge from the input image. After all side extensions are completed, we have a plus-sign-shaped region of known data shown on the top right of this figure. {\bf Bottom:} An example overlap-region used for corner extensions. We apply corner extensions after all four side extensions have been completed and we have a plus-sign-shaped region of known pixels, with four missing corners. To fill in these missing corners, we take an L-shaped region containing a set of both rows and columns of known pixel values along the edge of the missing corner. Similarly to the side extensions, we use an appropriately placed L-shaped overlap-region within the extension cells and require that this overlap-region has the same pixel values as the L-shaped region taken from the known pixel values.}
    \label{fig: Figure 3}
\end{figure}

Figure~\ref{fig: Figure 3} shows the different overlap-regions we use to extend the input images. We group these overlap-regions into two shapes: strips for side extensions and L-shapes for corner extensions. For each shape, we use four distinct overlap-regions, each associated with one of the four sides or four corners.
\begin{itemize}
    \item {\bf Side Extensions (Strips)}: We first extend each side of the input images. To extend on a given side, we require that the extension cells match a set of rows or columns on the edge of the input images. The overlap-region $\C$ is then a horizontal or vertical strip containing a set of rows or columns on the opposite edge of the extension cell (as illustrated in Fig.~\ref{fig: Figure 2}). After extending each side as illustrated in Fig.~\ref{fig: Figure 3}, top, we have a plus-sign-shaped spatial domain with known pixel values.

    \item {\bf Corner Extensions (L-shapes)}: Once all sides of the input images have been extended, there are still missing corners between each pair of adjacent sides (as shown in Fig.~\ref{fig: Figure 3}, top). To fill in these missing corners, we create additional corner extensions using new extension cells with differently shaped overlap-regions. For each of these corner extensions, we use an L-shaped overlap-region containing a set of both rows and columns within the extension cells. That is, we require that this L-shaped overlap-region within the extension cells matches a corresponding L-shaped region of known pixel values bordering the missing corner (as shown in Fig.~\ref{fig: Figure 3}, bottom). Figure~\ref{fig: Figure 3}, bottom shows the L-shape associated with the top-right missing corner; the remaining L-shapes are chosen similarly. Note that this overlap-region corresponds to pixel values generated by the previous side extensions (in addition to pixel values from the input images), so this algorithm is recursive: we generate new extension cells based on previous extensions.
\end{itemize}
Once all four sides have been extended and all four corners have been filled in at each time-step, we have a time-series of larger images (with the input images centered).

Recursive application of both the side and corner extensions can extend to an arbitrarily large $H_\ext \times W_\ext $ size. For $H_\ext \times W_\ext $ values much larger than $H\times W$, this algorithm requires multiple applications of ReVAR-FFBS for each of the eight unique overlap-regions. In these cases, ReVAR-ext first recursively applies the side extensions: for each subsequent extension, we extract new overlap-pixels from the edge of the most recently stitched extension cell (similarly to Fig.~\ref{fig: Figure 2}) and then stitch the new extension cell to the most recently generated cell. After the appropriate number of side-extensions, we fill in the missing corners using a similar recursive strategy.

\section{DATA AND METRICS}
\label{s: Data and Metrics}
In this section, we describe the data sets used to generate the results in this paper, along with our quality metrics.

\subsection{Measured Data Sets}
\label{s: Measured Data Sets}
We apply ReVAR-ext to two measured data sets \cite{Kemnetz_Data} from a wind tunnel experiment\cite{Kemnetz}. These data sets contain optical path difference (OPD) values, which are proportional to phase aberrations for a given wavelength \cite{Vogel}. Additional details about this experiment can be found in Refs.~\citenum{Kemnetz, KemnetzDissertation} and Sec. 5.2 of Ref.~\citenum{UtleyBoiling2}.

Table~\ref{tab: Experimental Data Sets} lists details of the OPD images resulting from this experiment. Here, the OPD data has units of $\mu$m. For each data set, we restricted to the largest square that can be inscribed in the images. Pre-processing for these measured data sets is discussed in Appendix C of Ref.~\citenum{UtleyBoiling2}.

\begin{table}[htbp]
    \caption{Measured Data Sets F06 and F12}
    \label{tab: Experimental Data Sets}
    \centering
    \setlength{\tabcolsep}{2em} 
    \begin{tabular}{l c c}
        \toprule
        \multirow{2}{*}{\textbf{Property}} & \multicolumn{2}{c}{\textbf{Data Set}} \\
        \cmidrule(lr){2-3}
         & \textbf{F06} & \textbf{F12} \\[0.5ex] 
        \midrule
        Image Size (pixels) & (22, 22) & (18, 18) \\
        Number of Time-Steps, $N_T$ & 150,600 & 251,100 \\
        Sampling Frequency, $f_s$ [kHz] & 100 & 130 \\
        \bottomrule 
    \end{tabular}
\end{table}

\subsection{Synthetic Data Sets}\label{s: Synthetic Data Sets}
In addition to the measured data sets F06 and F12, we also use ReVAR-ext to spatially extend synthetic data sets generated by ReVAR.

Here, we use pre-trained ReVAR models to generate synthetic data sets with the same statistics as the measured data sets F06 and F12 (see Ref.~\citenum{Utley2} for further details). Since ReVAR can generate time-series data of arbitrary duration, we use these synthetic data sets to evaluate ReVAR-ext on data with many time-steps. We denote these synthetic data sets F06-syn and F12-syn.

We generated five seconds of synthetic data for both F06-syn and F12-syn, where the images from each synthetic data set have the same size as those from the measured data sets (listed in Table~\ref{tab: Experimental Data Sets}). Using the sampling frequencies in Table~\ref{tab: Experimental Data Sets}, this amounted to 500,000 time-steps for F06-syn and 650,000 time-steps for F12-syn.

\subsection{Quality Metrics}
\label{s: Quality Metrics}
We assess ReVAR-ext using three quantities: the TPS of the OPD images (denoted $S_{OPD}$), the TPS of the streamwise slopes of the OPD images (denoted $S_{\theta_x}$), and the normalized spatial autocorrelation (AC) of the OPD images (denoted $R_{OPD}$).

To compare the TPS of input data with that of the extended data, we evaluate the extended data's TPS using only the newly generated pixels outside of the input data's image size. For details on our computation of the TPS from data, see Sec. 3.2 of Ref.~\citenum{Utley2} and Sec. 5.3 of Ref.~\citenum{UtleyBoiling}.

We compute the spatial AC from the input data by applying a biased AC estimator to each image in the time-series, averaging over time, and then normalizing by the total spatial variance. To compare the spatial AC of the input data with that of the extended images, we restrict the extended images' spatial AC to the spatial lags contained within the input data's spatial AC. We accomplish this by partitioning each extended image into a grid of smaller images that each have the same size as the input images. We use the above methodology to compute the spatial AC of each smaller image in this grid and then average over this grid.

We compute the error between the TPS and spatial AC of the input data and those of the extended images. Specifically, we take the normalized root mean square error (NRMSE), which we denote NRMSE$(\hat{\bm{y}}, \bm{y}^{\text{data}})$, where $\hat{\bm{y}}$ is computed from the extended images and $\bm{y}^{\text{data}}$ is computed from the input data. For more details, see Sec. 5.3 of Ref.~\citenum{UtleyBoiling2}.

Using the above notation, NRMSE$(\hat{\bm{y}}, \bm{y}^{\text{data}})$, we report the following errors:
\begin{itemize}
    \item \textbf{Slopes TPS Error}: NRMSE$(\hat{S}_{\theta_x}, S_{\theta_x}^{\text{data}})$. 

    \item \textbf{OPD TPS Error}: NRMSE$(\hat{S}_{OPD}, S_{OPD}^{\text{data}})$.

    \item \textbf{Spatial AC Error}: NRMSE$\left(\hat{R}_{OPD}, R_{OPD}^{\text{data}}\right)$.  
\end{itemize}

\section{RESULTS}
\label{s: Results}
In this section, we show results from ReVAR-ext applied to both the measured data sets F06 and F12 and the corresponding synthetic data sets F06-syn and F12-syn. For each data set, we extended the images to twice the height and twice the width of the input images.

\subsection{Parameter Estimation}\label{s: Parameter Estimation}
We applied ReVAR's parameter estimation algorithm described in Sec. 2.A of Ref.~\citenum{Utley2} to both measured data sets F06 and F12. For both measured data sets, we used four time-lags and two low-pass filters for ReVAR's linear autoregressive model (see Ref.~\citenum{Utley2} for further details) and estimated ReVAR's parameters from the first 80\% of the time-series; we set aside the remaining 20\% to use as input to ReVAR-ext. 

We derived the parameters of ReVAR-ext from the previously estimated parameters of ReVAR for both F06 and F12. We used these same derived parameters for extending both the measured and synthetic data sets.

\subsection{Results from Measured Data}\label{s: Results from Measured Data}

\begin{figure}[h]
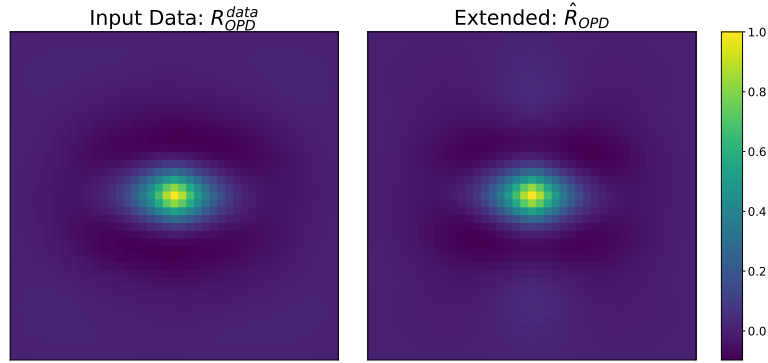

    \centering
    \includegraphics[width=0.7\textwidth]{F06_tps_plots_extend_only.pdf}

    \includegraphics[width=0.6\textwidth]{F06_ac_images_extend_only.pdf}
    \caption{ReVAR-ext applied to measured data set F06. {\bf Top:} Comparisons of the temporal power spectra obtained from measured data set F06 (blue) and from the spatial extension generated by ReVAR-ext (orange). The extended TPS matches the reference TPS for both slopes, $S_{\theta_x}$, and OPD, $S_{OPD}$, at all frequencies. Here, the OPD has units of $\mu$m. {\bf Bottom:} Comparison of the normalized spatial autocorrelation obtained from measured data set F06 (left) and from the extended images (right). The extended autocorrelation approximately matches the reference autocorrelation, $R_{OPD}$, but has inaccuracies at larger lags and at vertical lags.}
    \label{fig: F06 - Measured Data Plots}
\end{figure}

\begin{figure}[h]
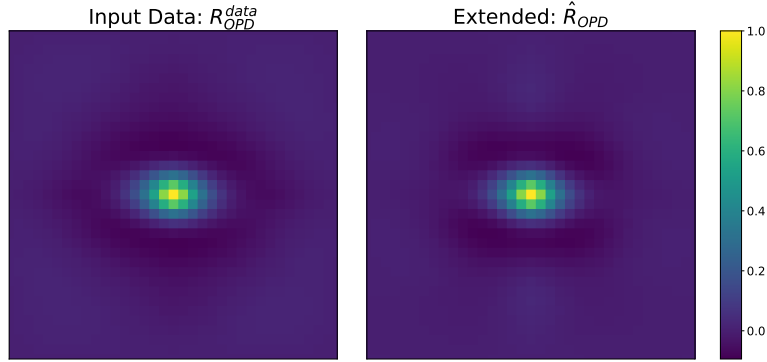

    \centering
    \includegraphics[width=0.7\textwidth]{F12_tps_plots_extend_only.pdf}

    \includegraphics[width=0.6\textwidth]{F12_ac_images_extend_only.pdf}
    \caption{Results analogous to Fig.~\ref{fig: F06 - Measured Data Plots} but for measured data set F12.}
    \label{fig: F12 - Measured Data Plots}
\end{figure}

In this section, we show results from extending the measured data sets F06 and F12. For both data sets, we applied ReVAR-ext to the remaining 20\% of the measured data. We generated ten independent extended data sets, each using the same 20\% input data. We computed the errors from Sec.~\ref{s: Quality Metrics} between each generated data set and the 20\% input data, then averaged over the ten data sets.

Figures~\ref{fig: F06 - Measured Data Plots} and~\ref{fig: F12 - Measured Data Plots} show the TPS and spatial AC results for measured data sets F06 and F12, respectively. These results show that ReVAR-ext closely matches the TPS and approximately matches the spatial AC for both data sets. While the extended AC closely matches the measured AC at small spatial lags, there are inaccuracies at larger lags and especially at vertical lags.

Table~\ref{tab: Error Metrics - measured data} shows the error metrics of ReVAR-ext for both measured data sets F06 and F12. As is consistent with the TPS plots in Figs.~\ref{fig: F06 - Measured Data Plots} and~\ref{fig: F12 - Measured Data Plots}, both the slopes and \textit{OPD} TPS errors are within $\approx$4\%. However, the spatial AC errors are much higher, ranging from $\approx$10-17\%. This suggests that ReVAR-ext closely matches temporal correlations, but only approximately matches spatial correlations.

\begin{table}[h]
    \caption{Error Metrics - ReVAR-ext Applied to Measured Data}
    \label{tab: Error Metrics - measured data}
    \centering
    \begin{tabular}{l c c c c}
        \toprule
        \multirow{2}{*}{\textbf{NRMSE (\%)}} & \multicolumn{2}{c}{\textbf{Measured Data Set}} \\
        \cmidrule(lr){2-3}
         & \shortstack{F06} & \shortstack{F12} \\[0.5ex] 
        \midrule
        Slopes TPS Error (\%) & \phantom{0}4.07 & \phantom{0}2.29 \\
        OPD TPS Error (\%) & \phantom{0}2.46 & \phantom{0}1.61 \\
        Spatial AC Error (\%) & 10.31 & 16.98 \\
        \bottomrule
    \end{tabular}
\end{table}

\subsection{Results from Synthetic Data}\label{s: Results from Synthetic Data}

\begin{figure}[h]
    \centering
    \includegraphics[width=0.7\textwidth]{F06_tps_plots_synthetic.pdf}

    \includegraphics[width=0.6\textwidth]{F06_ac_images_synthetic.pdf}
    \caption{ReVAR-ext applied to synthetic data set F06-syn. {\bf Top:} Comparisons of the temporal power spectra obtained from synthetic data set F06-syn (blue) and from the spatial extension generated by ReVAR-ext (orange). The extended TPS matches the reference TPS for both slopes, $S_{\theta_x}$, and OPD, $S_{OPD}$, at all frequencies. Here, the OPD has units of $\mu$m. {\bf Bottom:} Comparison of the normalized spatial autocorrelation obtained from synthetic data set F06-syn (left) and from the extended images (right). The extended autocorrelation approximately matches the reference autocorrelation, $R_{OPD}$, but has inaccuracies at larger lags and at vertical lags.}
    \label{fig: F06 - Synthetic Data Plots}
\end{figure}

\begin{figure}[h]
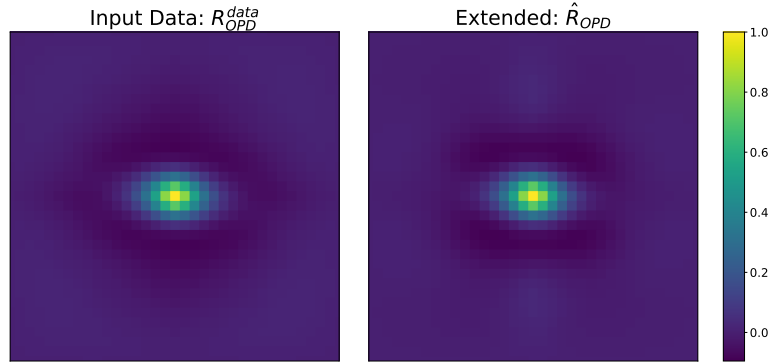

    \centering
    \includegraphics[width=0.7\textwidth]{F12_tps_plots_synthetic.pdf}

    \includegraphics[width=0.6\textwidth]{F12_ac_images_synthetic.pdf}
    \caption{Results analogous to Fig.~\ref{fig: F06 - Synthetic Data Plots} but for synthetic data set F12-syn.}
    \label{fig: F12 - Synthetic Data Plots}
\end{figure}

In this section, we show results from applying ReVAR-ext to the synthetic data sets F06-syn and F12-syn generated by ReVAR. For both synthetic data sets, we used the full synthetic time-series as input to ReVAR-ext. We generated ten independent extended data sets, each using the same input synthetic data. We computed the errors from Sec.~\ref{s: Quality Metrics} between each generated data set and the input synthetic data, then averaged over the ten data sets.

Figures~\ref{fig: F06 - Synthetic Data Plots} and~\ref{fig: F12 - Synthetic Data Plots} show the TPS and spatial AC results for synthetic data sets F06-syn and F12-syn, respectively. These results are largely consistent with the results from the measured data sets F06 and F12. ReVAR-ext closely matches the TPS and approximately matches the spatial AC of both synthetic data sets.

Table~\ref{tab: Error Metrics - synthetic data} shows the error metrics of ReVAR-ext for both synthetic data sets F06-syn and F12-syn. As with the plots in Figs.~\ref{fig: F06 - Synthetic Data Plots} and~\ref{fig: F12 - Synthetic Data Plots}, these errors are largely consistent with the results from measured data. The TPS errors are slightly smaller for these synthetic data sets, ranging from $\approx$1-3\%. However, the spatial AC errors are slightly higher for synthetic data when compared to the measured data, ranging from 11-18\%. This further establishes that ReVAR-ext closely matches temporal correlations but only approximately matches spatial correlations.

\begin{table}[h]
    \caption{Error Metrics - ReVAR-ext Applied to Synthetic Data}
    \label{tab: Error Metrics - synthetic data}
    \centering
    \begin{tabular}{l c c c c}
        \toprule
        \multirow{2}{*}{\textbf{NRMSE (\%)}} & \multicolumn{2}{c}{\textbf{Synthetic Data Set}} \\
        \cmidrule(lr){2-3}
         & \shortstack{F06-syn} & \shortstack{F12-syn} \\[0.5ex] 
        \midrule
        Slopes TPS Error (\%) & \phantom{0}1.19  & \phantom{0}1.55 \\
        OPD TPS Error (\%) & \phantom{0}2.15 & \phantom{0}0.95 \\
        Spatial AC Error (\%) & 11.15 & 17.43 \\
        \bottomrule
    \end{tabular}
\end{table}

\section{CONCLUSIONS}
\label{s: Conclusions}
In this paper, we introduced ReVAR-ext (Re-whitened Vector AutoRegression-extender), an algorithm for spatially extending measured aero-optic phase screens outside of their limited spatial domain. This algorithm builds on top of the existing ReVAR algorithm\cite{Utley2}, which generates arbitrary-duration time-series of aero-optic phase screens that closely match the spatial and temporal correlations of measured aero-optic data. ReVAR-ext takes in as input a time-series of fixed-sized phase screens and then extends each phase screen to an  arbitrarily large size. When paired with ReVAR, the ReVAR-ext algorithm can generate a time-series of synthetic aero-optic phase screens with arbitrary temporal duration and arbitrary spatial size.

We evaluated ReVAR-ext on both measured aero-optic data sets and synthetic aero-optic data sets generated by ReVAR. Our experiments show that the extended aero-optic phase screens generated by ReVAR-ext closely match the temporal power spectrum (TPS) of both the measured and synthetic data sets, with NRMSE ranging from $\approx$1-4\%. Further, ReVAR-ext approximately matches the spatial AC of the measured and synthetic data, with NRMSE ranging from 10-18\%. At larger spatial lags and especially at vertical lags, there is a mismatch between the spatial AC of the measured data and that of the extended data. Thus, ReVAR-ext closely matches the temporal correlations and approximately matches the spatial correlations of both measured and synthetic aero-optic phase screens.

\acknowledgments
C.A.B. was partially supported by the Showalter Trust.  J.W.U., G.T.B., and C.A.B. were partially supported by AFRL/RDKL FA9451-20-2-0008. The authors would like to thank the Showalter family and the United States Air Force for supporting this research.

\section*{DISCLAIMER}
The views expressed are those of the author and do not necessarily reflect the official policy or position of the Department of the Air Force, the Department of Defense, or the U.S. government. Approved for public release; distribution is unlimited.  Public Affairs release approval \# \PAnumber.  

\bibliography{report}
\bibliographystyle{spiebib}

\end{document}